\documentclass[conference,a4paper,final,twocolumn,10pt,twoside]{IEEEtran}
\usepackage{layout}
\IEEEoverridecommandlockouts 

\ifCLASSINFOpdf

\else
\fi
\usepackage{mathtools}
\usepackage{amsmath}
\usepackage{subcaption}
\usepackage{amssymb}
\usepackage{cite}
\usepackage{graphicx}
\usepackage{mathrsfs}
\usepackage{siunitx}
\usepackage{color}

\DeclareMathAlphabet{\mathscrbf}{OMS}{mdugm}{b}{n}
\usepackage{pdfpages}
\usepackage{graphicx}
\makeatletter
\renewcommand*\env@matrix[1][*\c@MaxMatrixCols c]{%
  \hskip -\arraycolsep
  \let\@ifnextchar\new@ifnextchar
  \array{#1}}
\makeatother
\newcolumntype{M}[1]{>{\centering\arraybackslash}m{#1}}
\newcolumntype{N}{@{}m{0pt}@{}}

\title{Performance Enhancement of the Golden Code \\ by Utilizing ORIOL Antenna}
\author{\IEEEauthorblockN{Vahid Amiri$^1 $, Mohammadali Sadat Hosseini$^{2}$, Ali Lotfi-Rezaabad$^3$, and Siamak Talebi$^{4,5}$ \IEEEmembership{, Member,~IEEE}}
\IEEEauthorblockA{$^1$Iran University of Science and Technology (IUST), Tehran, IRAN \\
$^2$Petroleum University of Technology (PUT), Tehran, IRAN\\
$^3$Sharif University of Technology (SUT), Tehran, IRAN\\
$^4$Shahid Bahonar University of Kerman (SBUK), Kerman, IRAN\\
$^5$Advanced Communication Research Instite (ACRI), Sharif University of Technology (SUT), Tehran, IRAN}\\
  
Email: v\_amiri@elec.iust.edu, Lotfi-rezaabad\_Ali@ee.sharif.edu, Siamak.Talebi@uk.ac.ir}

\begin{document}
\maketitle

\begin{abstract}
In this paper, a novel method is exposed to improve the performance of the Golden code, by using octagonal reconfigurable isolated orthogonal element (ORIOL) antennas, instead of a conventional microstrip patch antenna. The aforementioned antenna, should be employed in both the transmitter and the receiver sides, to approach the mentioned improvement. As a matter fact, in this paper, we recommend space-time-polarization diversity instead of space-time singly; therefore it is obvious that by employing the aforementioned technique, the system obtains more strength against destructive fading. The simulations for different rates have confirmed that, utilizing ORIOL antenna outperforms patch microstrip one, which is roughly about $2$ to $3$ dB, according to the rates. 
\end{abstract}
\section{Introduction}
\IEEEPARstart{N}{owadays}, on the one side, tend to have more wireless data rate is interminable, on the other side, spectrum is the most scare assets in the wireless communication systems \cite{Cisco}. Generally, customers favor wireless communication systems more often for its reasonable cost as well as its resilience. Moreover, high data rate in communication systems, causes more inter symbols interference (ISI) phenomenon. Most of modern communication systems such as long term elevation (LTE), utilize multi-input multi-output (MIMO) technique, which not only is a useful approach for digital narrow band wireless communication systems, but also affords more reliability for them \cite{5490976}. Lately, research area which attracts researchers' attention named as Massive MIMO, also forgives more credentials to MIMO systems \cite{6736761}.
 
Briefly, the main idea of MIMO technique is to transmit and receive signals from multiple antennas. Developing more efficient MIMO systems is a remain challenging, for instance, intelligent combining of symbols at one end and dissociating them at the other end is still summoning to contest. One of the most effective and practical methods against fading of the channel in the category of MIMO technique, is to utilize space time block codes (STBCs) \cite{jafarkhani2005space}. In this approach, on the transmitter side, symbols are transmitted by two or more antennas, in different time slots; while on the receiver side, signals are detected by two or more antennas, again in different time slots. In the other words, Space-time coding takes advantage of the spatial, as well the temporal diversities to combat the fading phenomenon in digital wireless communication. Alamouti's scheme \cite{730453}, is the most well-known STBC approach. In fact, the proposed code is both full rate and full diversity one. Moreover, this code owes its dominance to its simple decoder. Another well-known approach named as the Golden code. The Golden code was introduced in \cite{1412035}; the same as the Alamouti code, it is both full rate and full diversity code, however its rate is 2, which is the double of the Alamouti code. It is worthwhile to mention that the complexity of the Golden code's decoder is much more than the Alamouti one.                      

As it was mentioned before, one of the most critical issues of the LTE systems, is the ISI phenomenon. One way to combat ISI and fading phenomena as well, is to utilize diversity technique. In the other words, diversity means, creating copies of the same signal with uncorrelated fading coefficients are provided at the receiver. Hereupon, employing some of these techniques such as polarization, space, time, and frequency diversities can revamp the functionality of wireless system fiercely \cite{proakis1994communication}. 
Regard that, The maximum spatial diversity that STBCs can achieve, is not more than of the total number of antennas on the transmitter side \cite{jafarkhani2005space}.

There are several studies, in which the performance of MIMO systems with reconfigurable antenna have been investigated. Here, we review some of their dominant. In \cite{4011911}, reconfigurable was introduced for MIMO wireless systems. They claimed that by this amalgamation, additional degrees of freedom (DoF) are a rational expectation. In addition, they have defined the spatial correlation, to model the effect of reconfigurable antenna mismatch. In \cite{4463909}, authors  theoretically scrutinized the obtained capacity, by utilizing reconfigurable antenna. They also make their investigation more valuable through practical implementation. In another approach \cite{6060888}, authors recommend reconfigurable U-slot antenna for MIMO-OFDM wireless systems, to achieve more diversity gain.

Here, it is prerequisite to apprise some principal features of the reconfigurable antennas. Indeed, a reconfigurable antenna \cite{4463909}, is an antenna in which, principal parameters such as frequency and polarization can be controllable, even when they are radiating. In addition, the active devices which are utilized in such antennas, make the capability to adapt the radiation properties like bandwidth, radiation pattern, etc. The ORIOL reconfigurable antenna \cite{5313931}, is a two feed octagonal shaped microstrip patch antenna. There are four micro-electro-mechanical (MEM) switches in the feeds of the aforementioned antenna, that grant it the advantage of actualizing two linearly polarized electromagnetic waves. Proper planning and feeding of these switches, gives the antenna the chance to reconfigure its parameter efficiently. This antenna has two radiation states; vertical/horizontal state ($0^{\circ}/90^{\circ}, \text{or }\hat{x}/\hat{y})$, and slant state ($\pm45^{\circ} \text{or}~ (\frac{\hat{x}+\hat{y}}{\sqrt{2}})/(\frac{\hat{x}-\hat{y}}{\sqrt{2}}))$.

In this article, first, we express the significant features of the ORIOL antennas in detail. In the second step, we will clarify the Golden code. This is followed by the elucidation for amalgamation of the ORIOL antenna and the Golden code. The main contribution of this paper is to investigate the best possible combination of the ORIOL antenna, and the Golden code to take the most advantage of them. We also have scrutinized the proposed structure for various data rates, which causes to have more comprehensive conclusion, about the obtained enhancement. Finally, we discuss the simulation results and also make a performance comparison between the proposed structure and the traditional one. As it is a rational exception, our proposed structure outperforms the traditional one. 
 
The rest of the paper is organized as follows; In section II which is about the system model, the ORIOL antenna’s structure, fabrication, radiation properties, and the Golden code are expressed respectively. Section III, contains our contribution. Simulation results are provided in Section IV. Finally, Section V concludes the paper.

\textit{Notation}: We used bold letter for matrices. Superscripts $(.)^H, ||.||_F,$ and $(.)^*$ to indicate Hermitian, Frobenius norm and
complex conjugation, respectively. In addition, we use 
$\mathbb{C}^{M\times N}$ to represent the set of $M\times N$ matrices over field of complex numbers.

\section{System Model}
In this section, first we express the structure of the ORIOL antenna. This is followed by specification of the Golden code and other  prerequisite commentary.
\subsection{The ORIOL Antenna}
In the reconfigurable antennas, there are some active devices, in order to adjust and optimize the radiation properties. The main idea of utilizing this type of antennas, is that, they are used, to alter the propagation channel state, which enclosed by the transmitter and receiver. As a result, it will cause to enhance the performance of the MIMO communication systems. The ORIOL antenna is one of these antennas, that can operate in both pattern and polarization reconfiguration cases. This antenna works in two states; we use index $p$ for enumerating these states, and $n$ for enumerating feeding ports.

\begin{figure}
\centering
\includegraphics[trim={2cm 8cm 2cm 8cm},width=8cm,clip]{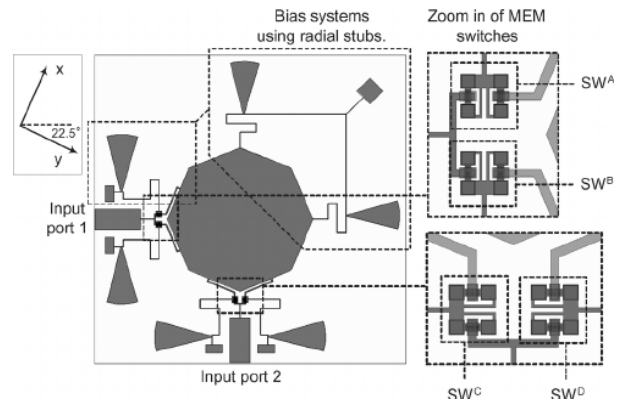}\\
\caption{schema of the ORIOL antenna.}
\label{ORIOLanteann}
\end{figure}

Owing to the fabrication of this antenna in \cite{5313931}, it is consists of Quartz substrate, with the relative permittivity $\epsilon_r=3.78$, the dissipation factor $\tan \delta=0.0002$, and its height is $1.575$ mm  as well. The perfect electric conductor (PEC), which acts as the radiating part of this antenna is made of gold. It is worthwhile to mention that antenna's bias electronic network is also made of gold with thickness of $t=0.5 ~\mu\text{m}$. Hereupon, the assembled antenna become expensive, nevertheless, there is a practical way, that gives us the power to mitigate the implementation's cost, which will explain more in the future work's section. To have better view of the ORIOL antenna, its schema is depicted in Fig. \ref{ORIOLanteann}. In two following subsections, the scattering parameters, and radiation pattern of the ORIOL antenna are briefly expressed.
\subsubsection{The Scattering Parameters}
\begin{figure*}[t!]
    \centering
    \begin{subfigure}[t]{0.5\textwidth}
        \centering
        \includegraphics[trim={2cm 6.5cm 1.5cm 6.5cm},width=8cm,clip]{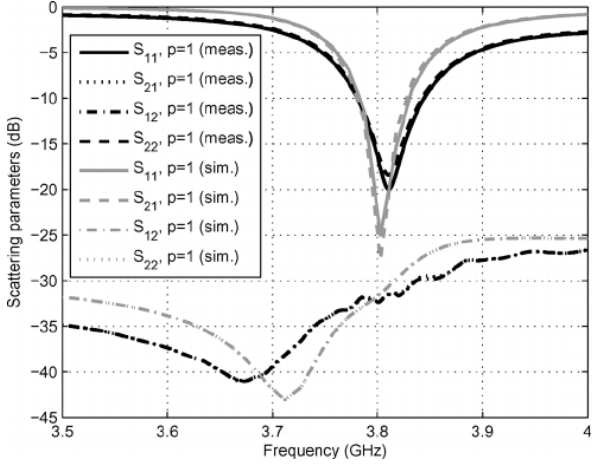}
        \caption{Scattering parameters on state $p=1$,}
        \label{scatter1}
    \end{subfigure}%
    ~ 
    \begin{subfigure}[t]{0.5\textwidth}
        \centering
        \includegraphics[trim={2cm 6.5cm 1.5cm 6.5cm},width=8cm,clip]{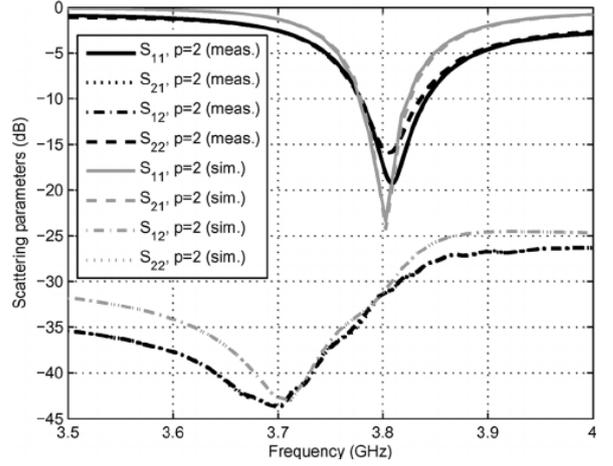}
        \caption{Scattering parameters on state $p=2$,}
        \label{scatter2}
    \end{subfigure}
    \caption{Measurement and simulation of the scattering parameters of the ORIOL antenna \cite{5313931}.}
\end{figure*}

The scattering parameters of this antenna have been shown in Fig. \ref{scatter1}, and Fig. \ref{scatter2} for different states of radiating. According to these figures, the resonant frequency of antenna is $3.82$ GHz, in addition, the return loss in each feeding ports, is near to $15$ dB, and the isolation between the two feeding ports, is $25$ to $45$ dB for both states. Moreover, $10$ dB bandwidth of this antenna is $64.9$ MHz ($1.7\%$ in $3.82$ GHz). Finally, the aforementioned values cause an extraordinary polarization purity for radiated electromagnetic waves.

\subsubsection{The Radiation Pattern}
\begin{figure}
\centering
\includegraphics[trim={1cm 4cm 1cm 4cm},width=8cm,clip]{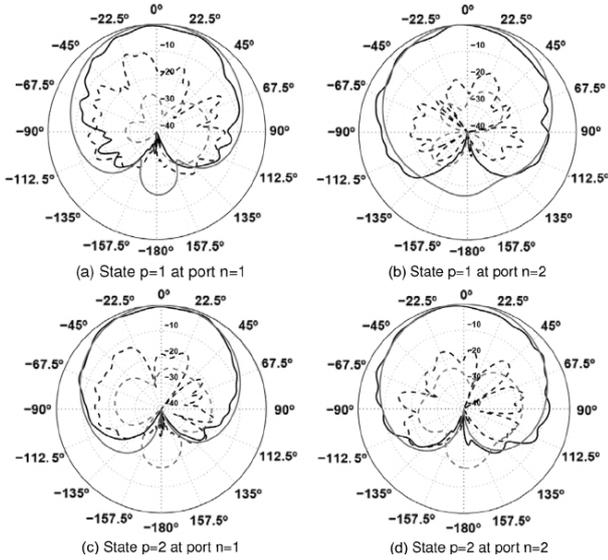}\\
\caption{Co (solid line) and cross(dash line) polarization of the normalized radiation patterns of the ORIOL antenna in the $E$ and $H$ planes, \cite{5313931}.}
\label{pattern}
\end{figure}

The far field pattern of this antenna for both states, $p=1, $ and $p=2, $ are depicted in Fig. \ref{pattern}. In the state $p=1$, the co-polar radiation at ports $n=1 (a), $ and $n=2 (b), $ are in the $\hat{y},$ and $\hat{x},$ directions, respectively. Likewise, in the state $p=2$, the co-polar radiation at ports $n=1 (c), $ and $ n=2 (d), $ are in the $(\frac{\hat{x}+\hat{y}}{\sqrt{2}}),$ and $(\frac{\hat{x}-\hat{y}}{\sqrt{2}}),$ directions, respectively.
 
\subsection{MIMO Wireless System}

In this subsection, first we will explain the channel model of space-time block codes and its decoder as well, after that, we explain the Golden code's structure, exclusively.

\subsubsection{Channel Model}
In this article, we deal with a quasi-static flat fading channel, with $M_T$ and $M_R$ antennas at the transmitter and the receiver, respectively. Assume that there is no knowledge about the channel state on the transmitter side, nonetheless, the receiver has informed about channel state information very well. Consider that through this paper, we have assumed that the transmitter and the receiver have been perfectly synchronized. Assume that the transmitter sends the codeword $\textbf{C}\in\mathbb{C}^{T\times M_T} $, over $T$ time slot, therefore we can model the MIMO wireless system as:

\begin{equation}\label{primaryequation}
\mathbf{Y} = \mathbf{CH}+\mathbf{N},
\end{equation}
where $\mathbf{Y}\in\mathbb{C}^{T\times M_R}$, is the received signal, $\mathbf{H}\in\mathbb{C}^{M_T\times M_R},$ is fading
matrix, and $\mathbf{N}\in\mathbb{C}^{T\times M_R}$ is used to model  the Additive White Gaussian Noise (AWGN). For a system that has been designed for two blocks fading, the following expression can be expressed:

\begin{equation}\label{BlockFading}
\begin{bmatrix}
\mathbf{Y}_1 \\
\mathbf{Y}_2
\end{bmatrix}= 
\begin{bmatrix}
\mathbf{C}_1 & 0 \\ 0 & \mathbf{C}_2
\end{bmatrix}
\begin{bmatrix}
\mathbf{H}_1 \\ \mathbf{H}_2
\end{bmatrix}+
\begin{bmatrix}
\mathbf{N}_1 \\ \mathbf{N}_2
\end{bmatrix},
\end{equation}
in which the codeword can be defined as below:
\begin{equation}\label{CodeWord}
\mathbf{C}=\text{diag}\{\mathbf{C}_1,\mathbf{C}_2\}\in \mathbb{C}^{2T\times 2M_T}
\end{equation}

\subsection{Decoder Model}
To detect symbols with a maximum-likelihood (ML) detector at the receiver, the
decoder has to examine all possible answers for this equation, and
then decides on the minimum of the following equation:
\begin{equation}\label{decoder}
\hat{\mathbf{C}}=\arg\min_{\mathbf{C}^i}||\mathbf{Y}-\mathbf{C}^i\mathbf{H}||^2_F.
\end{equation}

\subsection{The Golden Code}
The Golden code is a full rate and full diversity code. Having neither  lower bound for its code gain distance (CGD), nor vanishing minimum determinant, are the most significant of its features, that make it dominance. Moreover, its minimum CGD is independent from the constellation size. In fact, the Golden code is the optimum space-time code for $2\times2$ MIMO systems, which is based on the golden number, $\frac{1\pm\sqrt{5}}{2}$; it is also has the best performance in comparison with the other codes in this category. Consider that the sequence $\mathscr{S}=\{S_1,S_2, \dots, S_M\},$ in which $S_i$ is drawn from the constellation such as BPSK or M-QAM. It is assumed that the symbols of sequence should be sent from the wireless channel. Owing to the Golden code, the codeword can be stated as follow:
\begin{equation*}
\mathbf{C}=\frac{1}{\sqrt{5}}
\begin{bmatrix}
\alpha(S_1 + \theta S_2) & \alpha(S_3 + \theta S_4) \\
\gamma\bar{\alpha}(S_3 + \bar{\theta}S_4) & \bar{\alpha}(S_1+\bar{\theta}S_2)
\end{bmatrix},
\end{equation*}
in which $\theta=\frac{1+\sqrt{5}}{2}$, $\theta=\frac{1-\sqrt{5}}{2}$, $\alpha=1+j(1+\theta)$, and $\alpha=1-j(1-\bar{\theta})$; in addition, $\gamma=e^{j\phi}$, where $\phi=\frac{\pi}{2}$. It is worthwhile to note that the elements of the mentioned codeword, may have both real and imaginary parts. However, beside its antecedent performance, it suffers from high complexity decoder. Nonetheless, some studies such as sphere decoder \cite{1468474}, are capable to mitigate such complexities.

\section{Proposed Method}
In this part,  we will propose our contribution, which is the amalgamation of the ORIOL antenna and the Golden code, in a well-organized manner to take the most advantage of them.  Nevertheless, some manipulations in the Golden code is inevitable. To define the designed codeword for the proposed structure, in the first step, we should define sub-codewords which are based on the Golden code.  To the best of our knowledge, the superlative method is to define sub-codewords $\mathbf{C_1}$ and $\mathbf{C_2}$ as follow: 
\begin{equation}{\label{c1}}
\mathbf{C_1}=
\begin{bmatrix}
\alpha(S_1 + \theta S_2) & \alpha(S_3 + \theta S_4)
\end{bmatrix},
\end{equation}
\begin{equation}{\label{c2}}
\mathbf{C_2}=
\begin{bmatrix}
\gamma\bar{\alpha}(S_3 + \bar{\theta}S_4) & \bar{\alpha}(S_1+\bar{\theta}S_2)
\end{bmatrix}.
\end{equation}
According to the Eq. \ref{CodeWord}, now we can define the codeword $\mathbf{C}_\text{ORIOL}$ for the proposed approach as Eq. {\ref{CORIOL}}.

\begin{table*}[t!]
\label{procedure}
\caption{Corresponds table between the ORIOL antenna raidation states, the sates of the MEMs switches, and the proposed codeword}
\begin{center}
\begin{tabular}{|M{2.5cm}||M{2.5cm}|M{2.5cm}|M{2.5cm}|M{2.5cm}|}
\hline
\hline
\multicolumn{5}{|c|}{\textbf{The transmission policy for the proposed system}}\\
\hline
\hline
\hline
port numbers &$n=1$ & $n=1$ & $n=2$& $n=2$ \\
\hline
state numbers &$p=1$ & $p=2$ & $p=1$& $p=2$ \\
\hline
$T=1$& $\alpha(S_1 + \theta S_2)$& off& $\alpha(S_3 + \theta S_4)$ & off\\
\hline
$T=2$& off & $\gamma\bar{\alpha}(S_3 + \bar{\theta}S_4)$& off & $\bar{\alpha}(S_1+\bar{\theta}S_2)$ \\
\hline
\hline
\end{tabular}
\end{center}
\end{table*}

\begin{figure*}[!t]
\normalsize
\setcounter{equation}{6}
\hrulefill
\begin{align}\label{CORIOL}
\mathbf{C}_\text{ORIOL}=\frac{1}{\sqrt{5}}
\begin{bmatrix}
\alpha(S_1 + \theta S_2) & \alpha(S_3 + \theta S_4) & 0 & 0\\
0 & 0 &  \gamma\bar{\alpha}(S_3 + \bar{\theta}S_4) & \bar{\alpha}(S_1+\bar{\theta}S_2)
\end{bmatrix}.
\end{align}
\hrulefill 
\end{figure*}
To ponder more on the functionality of the proposed system, we briefly elucidate it. Consider the codeword in Eq. \ref{CORIOL}, in the first time slot, $T=1$, the system acts as follows: It delivers $\alpha(S_1+\theta S_2)$ to the first port of the ORIOL antenna $(n=1)$; this procedure should be eventualized while the ORIOL antenna is being on the state one ($p=1$). Moreover, in the same time slot as previous, $\alpha(S_3 + \theta S_4)$ is also given to the second port of the ORIOL antenna which is the state one $(n=2)$. To accommodate the procedure, we provide Table. I. "off`` in the table is  indicating lack of action for the antenna.

It should be mentioned here that, why the system model employed in this paper might seem, somewhat different from the typical system model investigated by other papers on STBCs. As it can be seen in the codeword matrix, it seems to be four transmitting antennas are deployed on the transmitter side, however, it is not true.  Specifically, there are four virtual antennas.  In other words, we have arranged the codeword in a fashion which elaborates the ORIOL antenna to act in such a manner, that empower two of these four virtual antennas periodically.  Hence, in code word matrix, zero is used to indicate the out of work antennas.

\begin{figure}
\centering
\includegraphics[trim={8cm 3cm 6cm 3cm},width=10cm,clip]{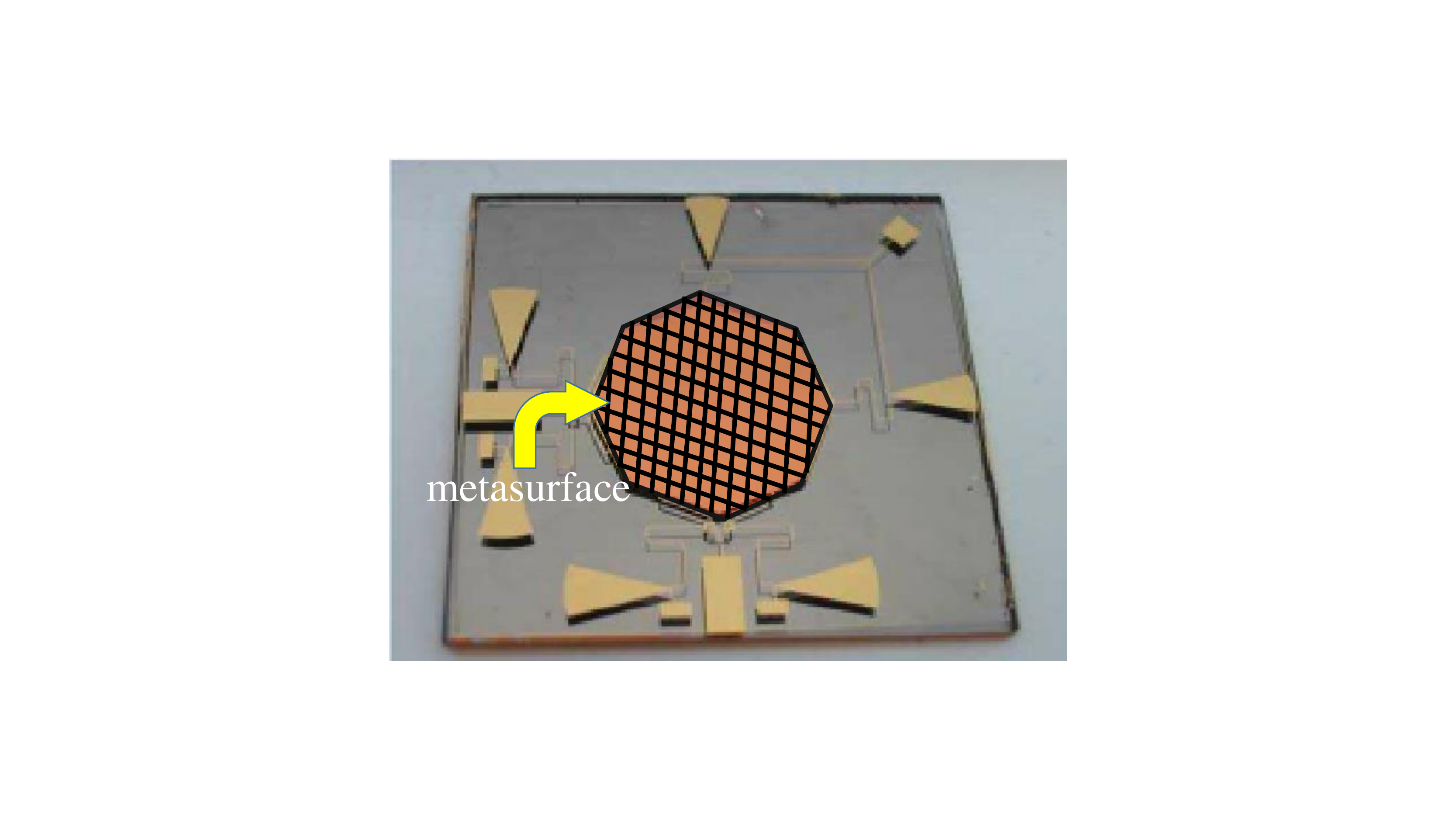}\\
\caption{An example of the location for the metasurface superstrate.}
\label{meta}
\end{figure}

One of the drawbacks of the ORIOL antenna is, its realized gain, which is about 4.8 dB, however, the suitable gain of this planar antenna is about $6$dB. In addition, the high cost is another issue with the ORIOL antenna.
Here, we recommend to utilize some extraordinary methods, which are based on metasurface and $3$D metamatrials, to overcome this phenomena. According to IEEE definitions \cite{6758443}, metamaterials and metasurfaces are the structures, that consist of sub-wavelength scatterers. A group of The afforementiond sub-wavelength scatters are able to create electromagnetic properties, that cannot be obtained naturally. One of the terrific and affordable ways to ameliorate both the radiation gain, and the bandwidth of this type of planar antennas (the ORIOL antenna), is to elaborate a metasurface superstrate above the radiating patch of the antenna, as well as metamaterial loading in the substrate of the antenna. Moreover, utilizing electromagnetic band gap (EBG) structures bellow, close to the radiating patch in antenna construction is feasible but, it is a bit feeble in comparison with metasurface and metamaterial. For instance, an utilization of a metasurface superstrate structures to enhance the radiation properties of the ORIOL antenna is represnted in Fig. {\ref{meta}}.

\section{Simulation Results}
\begin{figure}
\centering
\includegraphics[trim={4cm 8.8cm 4cm 8.5cm},width=8cm,clip]{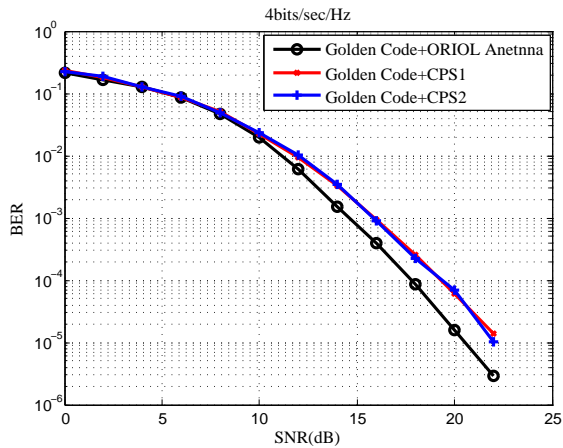}\\
\caption{Bit error rate versus signal to noise ratio for $4$QAM constellation
$4$ bit/sec/Hz.}
\label{4qam}
\end{figure}

\begin{figure}
\centering
\includegraphics[trim={4cm 8.8cm 4cm 8.5cm},width=8cm,clip]{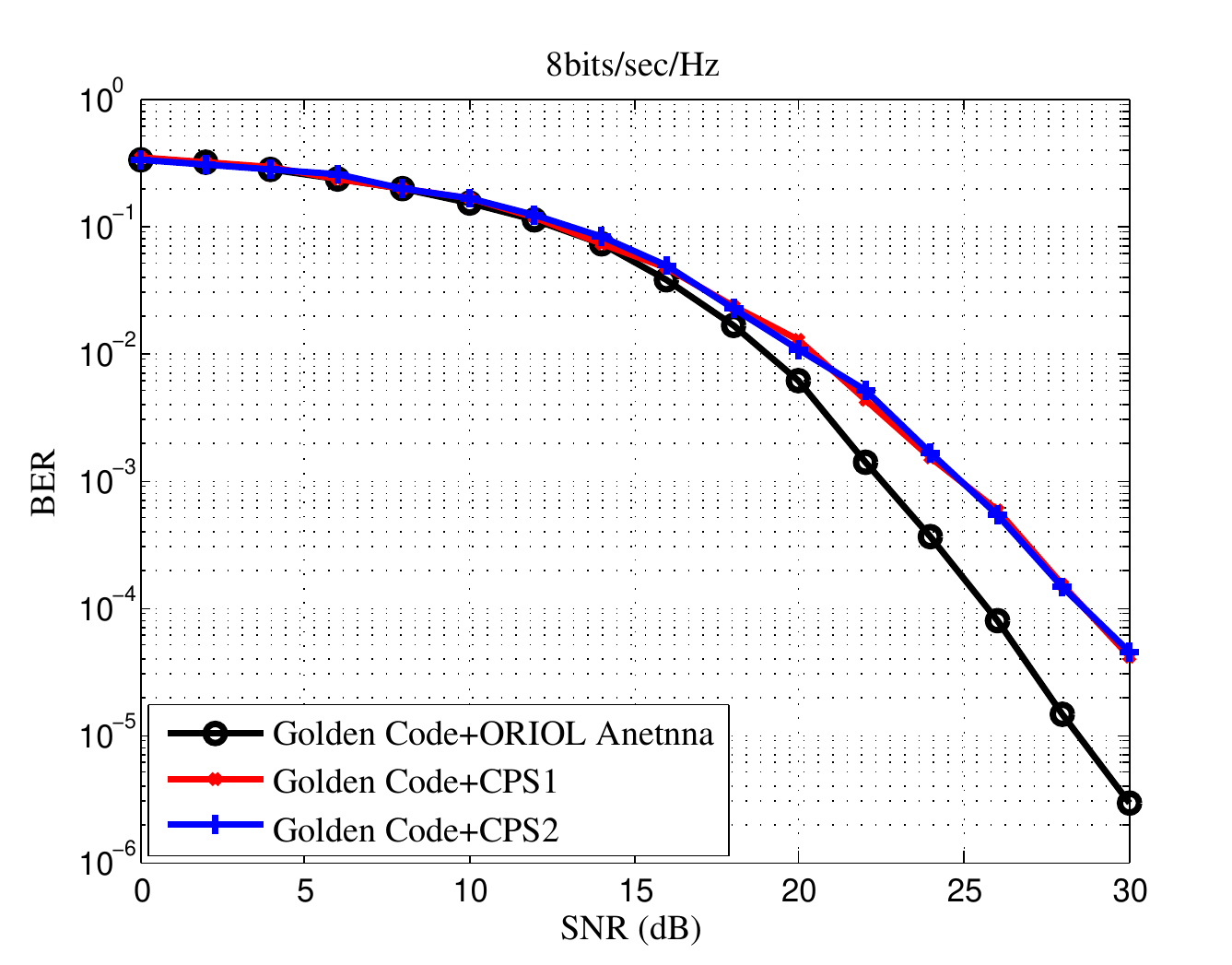}\\
\caption{Bit error rate versus signal to noise ratio for $16$QAM constellation
$8$ bit/sec/Hz.}
\label{16qam}
\end{figure}

Scrutinizing the performance of the proposed method, is provided in this section. To have a more holistic comparison performance, we also yield the performance of the popular approach named as the Golden code as well. The simulations were executed in an open loop MIMO communication system, using the sphere-decoding algorithm to detect symbols. After Monte-Carlo simulation, the final results are provided to study our approach with the Golden code. It is worthwhile to point out that, the simulations for our proposed method and the Golden code are done in a fair condition. In Fig. \ref{4qam}, we depicted bit error rate (BER) versus signal to noise ration (SNR). For that simulation, we utilize $4$QAM, which is one of the most popular constellations in digital communications. As it is undeniable, our approach surpasses the Golden code by an amount near to $2$dB for this constellation. Moreover, we also scrutinized both our approach and the Golden code for $16$QAM. The simulation results for the aforementioned constellation have been shown in the Fig. {\ref{16qam}}. Again, it is clear that our approach outperforms the Golden code close to $3$dB.

It is prerequisite to emphasize that, both the proposed structure and the Golden code have the same order of complexity $\mathscr{O}(m^4)$. The same order of complexity also approves that both of them are executed in the same and of course fair conditions. however, the sphere-decoding algorithm is an efficient method to lower the complexity of their decoder. 

\section{Conclusions}

In this article, we have expounded the ORIOL antenna comprehensively, in the first step. In the second one, we have described the Golden code in detail. Followed by them, we proposed an efficient method to leverage the ORIOL antenna for enhancing the Golden code's performance. However, it was obligated to modify the structure of the Golden code to reach the best possible realization, efficiently. Hereupon, we had been separating the codeword into two different sub-codewords. After that, we offered a planning for the ORIOL antenna. Following step by step of this procedure causes an efficient $2\times2$ STBC, which outperforms the optimum STBC (the Golden code), dramatically. This enhancement can benefit wireless communication in different ways, such as energy efficiency, which is the major part of green communications.
 
As well, we also propose some modernized methods, such as metasurfaced superstrate to not only mitigate the antenna fabrication's cost, but also modify propagation properties of the ORIOL antenna. 
\bibliographystyle{IEEEtran}

\bibliography{IEEEabrv}
\end{document}